\documentclass[12pt,a4paper]{article}
\usepackage{a4wide}
\usepackage{graphicx}

\usepackage{amsmath}
\usepackage{amssymb}
\usepackage{aas_macros}

\def\ll#1#2{\tilde{\lambda}_{#1}.\tilde{\lambda}_{#2}}

\def\vec#1{\boldsymbol{#1}}
\def\dd{\mathrm{d}}
\counterwithout{figure}{section}

\begin{document}
	\title{\large\bf Level order of quark systems:\\
		The puzzle of the Roper resonance, and related questions%
		\thanks{Submitted to the Fest Roper volume at Acta Physica Polonica B, edited by Michal Praszalowicz \& Igor Strakovsky}
	}
	\author{ Jean-Marc Richard\\
{\small Institut de Physique des 2 Infinis de Lyon, CNRS-IN2P3}\\
{\small 		Universit{\'e} Claude Bernard (Lyon 1) }\\
{\small 		4 rue Enrico Fermi, 69622 Villeurbanne, France}}
	\date{\small \today}
	\maketitle
	\begin{abstract}
		The problem of  ordering of radial vs.\ orbital excitations is reviewed. It is shown that the current quark models cannot explain the location of the Roper resonance which is slightly lower than the lowest negative-parity excitations. We also study some related spectral problems, such as the dependence of the energies  on the quark masses, and the possibility of bound states in simple chromelectric models. 
	\end{abstract}
\section{Introduction}
As far as I remember, I always used the results of David Roper, especially the ones dealing with pion-nucleon and nucleon-nucleon scattering. As a post-doc at Stony Brook, I was introduced to the pioneering remote access to data and analyzing programs set-up by the group of Virginia Tech, anticipating to a large extent the internet, with just a FTS phone-network and an acoustic interface between a phone and the local computer. 

Somewhat later, I had the chance to visit my colleague and friend Tetsuro Mizutani at Blacksburg and to meet there David Roper and Richard Arndt. With David I had a long discussion on the Roper resonance and its much debated location in the spectrum. Richard showed me his devices for computing the electromagnetic fields, drawing classical trajectories, estimating bound states and phase-shifts in quantum mechanics, etc.,  custom-designed for his students. This anticipated years ahead tools that are now offered by personal computers and their handy mathematical softwares. I was fascinated by the enthusiasm and innovative spirit of David and Richard. 

The famous Roper resonance \cite{Roper:1964zza} is the first excitation of the nucleon spectrum, with the same quantum numbers ${1/2}^+$ as the nucleon, a width of about 180\,MeV and a mass 1440\,MeV \emph{below} the orbital excitations at 1520\,MeV (${1/2}^-$) and 1535\,MeV (${3/2}^-$). Its low mass has always been found rather intriguing~\cite{RevModPhys.91.011003}. 
\section{The quark model of baryons}
The speculations on the composite nature of baryons started in the early 60s or even earlier, see e.g., \cite{Zweig:1980nu}, and a first convincing evidence was provided at the beginning of 1964 by the discovery at Brookhaven of the $\Omega^-$ baryon with strangeness $S=-3$~\cite{Barnes:1964pd}.

Explicit quark models were constructed, in particular by Morpurgo~\cite{Morpurgo:1965xy}, Greenberg~\cite{Greenberg:1964pe}, etc. The most comprehensive study was done by Dalitz \cite{1965hep..conf..251D}, and developed in the UK with his coworkers and emulators \cite{1977NuPhB.129...45J,Hey:1982aj}.
The harmonic oscillator (HO) model of Dalitz et al.\ was subsequently extended and popularized by Isgur and Karl \cite{Isgur:1983fs} and others, see, e.g.,  \cite{Gromes:1979xn,Forsyth:1982dq}. 

\begin{figure}[ht!]
	\centering
\includegraphics[width=0.4\linewidth]{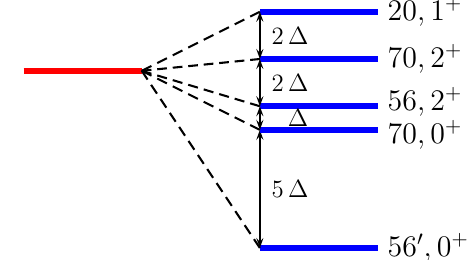}\quad
	\includegraphics[width=0.4\linewidth]{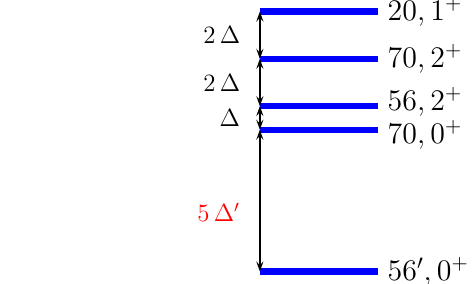}
	\caption{Left: Splitting of the $N=2$ level of the harmonic oscillator due to an anharmonic correction $v(r)$ of pairwise character treated to first order. Right:  $\Delta'\neq \Delta$ for the lowest splitting, if the perturbation contains a 3-body interaction.}
	\label{fig:figs-roper-fig1}
\end{figure}

The HO model, though extremely simple, accounts rather well for the main features of the baryon spectrum. Most discrepancies can be cured by various spin-independent and spin-dependent ``anharmonicity corrections'' that are spelled out in the above references. A problem however remains: the low mass of the first radial excitation, namely the Roper resonance.  In a pure HO model, the Roper comes above  the ground state twice higher than the first orbital excitation, while it is experimentally degenerate with the latter, or even slightly lighter. Isgur and Karl, Gromes and Stamatescu, Bowler and Tynemouth \cite{Isgur:1978wd,Gromes:1979xn,PhysRevD.27.662}, and others, following Dalitz et al.,  showed that if the HO pair potential $\propto r_{ij}^2$ is perturbed by some reasonable anharmonic potential $v(r_{ij})$, the $N=2$ level is split into five states with $[20,1^+]$ the highest and the Roper $[56,0^+]$ the lowest, as shown in Fig.~\ref{fig:figs-roper-fig1}. The notation, that looks a little cryptic today, associates the dimension of the SU(6) representation to the spin-parity $J^P$. Note that if the anharmonicity contains a 3-body potential, the pattern of splittings of the upper levels $[20,1^+]$, \dots $[70,0^+]$ is unchanged but the lowest splitting is different, with $\Delta'\neq\Delta$~\cite{Gromes:1979xn,PhysRevD.27.662}.  It could then be argued  that the Roper becomes close to the orbital excitation, though quantum mechanics textbooks tell us that first order perturbation, to be valid, requires the correction to be smaller than the separation of the unperturbed levels!

Indeed, any  exact 3-body calculation in a typical pairwise interquark potential shows that the Roper is obtained above the negative-parity excitation, see, e.g., \cite{Silvestre-Brac:1985aip}, and it can be checked that this result survives the admixture of a  a large class of spin-independent 3-body components.

In the 2-body case, there are rigorous results about such ordering \cite{Grosse:847188}. In particular, if the quark-antiquark potential  has a positive Laplacian, then $E(n,\ell)> E(n-1,\ell+1)$ if $E(n,\ell)$ denotes the energy of the level with $n$ nodes and orbital momentum $\ell$. 

In the 3-body case, an almost rigorous proof of the ordering was provided in \cite{Hogaasen:1982rb}. Consider the 3-quark Hamiltonian:
\begin{equation}\label{eq:H3}
	H=\sum_{i=1}^3\frac{\vec{p}_i^2}{2\,m}	+V(\vec r_1,\vec r_2,\vec r_3)~,
\end{equation}
where $V$ is symmetric and translation-invariant, not necessarily pairwise. As the color degree of freedom endorses the antisymmetry requirement and the spin-isospin wavefunction is symmetric, one can treat \eqref{eq:H3} as a model for three bosons. In the hyperspherical formalism, the ground state and its radial excitations can be expanded on the symmetric, scalar harmonics $\mathcal{P}_{[L]}$ as:
\begin{equation}\label{eq:exp-Psi}
	\Psi(\rho, \Omega_5)=\sum_{[L]} \frac{u_{[L]}(\rho)}{\rho^{5/2}} \mathcal{P}_{[L]}(\Omega_5)=
		\frac{u_0(\rho)}{\rho^{5/2}}\,\pi^{-3/2}+\frac{u_4(\rho)}{\rho^{5/2}}\,\mathcal{P}_4(\Omega_5)+\cdots~,
\end{equation}
where $[L]$ denotes the grand orbital momentum $L$ and its associated magnetic numbers. Here $\rho$ denotes the hyperradius given by $\rho^2=\vec x^2+\vec y^2$ in terms of the Jacobi coordinates $\vec x=\vec r_2-\vec r_1$ and $\vec y=(2 \, \vec r_3-\vec r_1-\vec r_2)/\sqrt3$, and the 5 angles of $\Omega_5$ include $\hat{\vec x}$, $\hat{\vec{y}}$, and $\tan^{-1}(|\vec y|/|\vec x|)$.  Similarly the potential energy $V$ can  itself be expanded into multipoles, say:
\begin{equation}\label{eq:exp-V}
V(\vec r_1,\vec r_2,\vec r_3)=V_0(\rho)+V_4(\rho) \,\mathcal{P}_4(\Omega_5)+\cdots 
\end{equation}
The remarkable feature in Eqs.~(\ref{eq:exp-Psi}-\ref{eq:exp-V}) is that the first correction to the hyperscalar approximation starts only at $L=4$, so that the single hyperradial equation:
\begin{equation}\label{eq:hyper-L0}
	-\frac{u''_0(\rho)}{m}+ \frac{15\,u_0(\rho)}{4\,m\,\rho^2} +V_0(\rho)\,u_0(\rho)=E_0\,u_0(\rho)~,
	\end{equation}
is an excellent approximation for the energy $E_{0,0}$ of the ground state and $E_{0,1}$ of  the Roper excitation. 

Similarly, the first orbital excitations (in $\hat{\vec x}$ or $\hat {\vec y}$) are well described in the approximation of the lowest harmonics
\begin{equation}\label{eq:orb-exc}
\Psi(\rho,\Omega_5)\frac{u_1(\rho)}{r^{5/2}}\,\mathcal{P}_1^{x,y}(\Omega_5)~,
\end{equation}
leading to the companion radial equation:
\begin{equation}\label{eq:hyper-L1}
		-\frac{u''_1(\rho)}{m}+ \frac{35}{4\,m\,\rho^2} u_1(\rho)+V_0(\rho)\,u_1(\rho)=E_1\,u_1(\rho)~,
\end{equation}
with, remarkably, the \emph{same} hyperradial potential $V_0(\rho)$ as in \eqref{eq:hyper-L0}. Moreover, this potential is obtained from the interquark potential by an averaging with positive weight that  keeps the sign of the Laplacian. In particular, a pair potential of the form $v(r)=-a/r+ b\, r$, with $a,\,b>0$ builds a hypercentral $V_0(\rho)=-A/\rho + B\,\rho$ with $A,\,B>0$, and a 3-body term with a positive Laplacian also results into $V_0(\rho)$ growing  faster than $-1/\rho$. Then the theorems on the level order derived for mesons with integer orbital momentum $\ell$ can be applied to Eqs.~(\ref{eq:hyper-L0}) and (\ref{eq:hyper-L1}) with angular momentum $\ell=3/2$ and $\ell=5/2$, respectively, i.e.,
\begin{equation}\label{eq:Roper-ord}
\Delta V>0 \ \Rightarrow\ 	E_{0,0}<E_{1,0}<E_{0,1}~,
\end{equation}
which expresses the Roper puzzle: in any local, symmetric potential growing faster than a Coulombic interaction, the first radial excitation $E_{0,1}$ is \emph{above} the orbital one $E_{1,0}$. 

Several ways-out have been explored, such as: relativistic kinematics, a spin-dependent interaction that acts differently on the orbital and radial excitation, a three-body force, 2-body and 3-body terms with components of negative Laplacian, non-local terms arising from the coupling to the decay channels, etc. See, e.g., \cite{Gavela:1978bz,Desplanques:1992rf,Wagenbrunn:2000ue}. However, the effects are often simultaneous in the models, and it is not clear which one is crucial for the proper level ordering.

\section{Splitting of other levels}
The splitting pattern of Fig.~\ref{fig:figs-roper-fig1} of the $N=2$ level has been  further studied and extended to other states. 

First of all,  it was observed that the spacing $\Delta=E[56,2^+]-E[70,0^+]$ is positive for any plausible potential. More precisely, it can be shown~\cite{Richard:1989ra} that:
\begin{equation}
	\frac{\dd}{\dd r}\genfrac{(}{)}{}{}{\dd v}{r\,\dd r}\gtrless 0
	\ \Rightarrow \ \Delta\gtrless 0~,
	\end{equation}
i.e., $\Delta>0$ if the pairwise perturbation $\sum v(r_{ij]})$ corresponds to $v(r)$ being a convex or concave function of $r^2$.

For $N=3$ and higher, analyzing the splitting patterns requires more and more sophisticated algebraic and group-theoretical tools
 \cite{Forsyth:1982dq,Richard:1989ra,Stancu:1991cz,Dmitrasinovic:2018irg}. Again the perturbation around a pure HO interaction can be generalized to perturbation around a purely hypercentral interaction. 
\section{Ordering as a function of the quark masses}
So far, we have discussed the spectrum for given constituent masses.  When one introduces in the model strangeness and heavy flavors, another question arises: how does the baryon mass evolve, when the constituent are changed? We shall restrict ourselves to the case where the potential energy does not depend on the quark masses, i.e., the case of flavor independence. 

The Hamiltonian reads:
\begin{equation}\label{eq:H3p}
H=\sum_{i=1}^3\frac{\vec p_i^2}{2\,m_i}+V(\vec r_1,\vec r_2,\vec r_3)~,
\end{equation}
where $V$ is translation-invariant and independent of the $m_i$. The ground-state energy is denoted $E(m_1,m_2,m_3)$. Obviously, the energy $E$ decreases, when one of the masses  $m_i$ increases, as the coefficient of $1/m_i$ is a positive operator. For instance:
\begin{equation}
	E(m_b,m,m)<E(m_c,m,m)<E(m_s,m,m)\quad \text{if}\quad m_s<m_c<m_b~.
\end{equation}
In the 2-body case, one can demonstrate that the energy is a concave function of the $1/m_i$, in particular \cite{Grosse:847188}
\begin{equation}\label{eq:BM}
2\,\mathcal{M}(m_1,m_2)\ge \mathcal{M}(m_1,m_2)+\mathcal{M}(m_2,m_2)~,
\end{equation}
for any flavor-independent quark-antiquark potential, where $\mathcal{M}(m_1,m_2)=E(m_1,m_2)+m_1+m_2$ is the mass of the meson made of the quark $m_1$ and antiquark $m_2$. 

It is tempting to generalize \eqref{eq:BM} to baryons as \cite{Nussinov:1983hb,Richard:1985cu}
\begin{equation}\label{eq:BM3}
	2\,\mathcal{M}(m_1,m_2,m\ge \mathcal{M}(m_1,m_2,m)+\mathcal{M}(m_2,m_2,m)~,
\end{equation}
which, indeed, turns out to be true for any reasonable interquark potential. However, as shown by Lieb \cite{Lieb:1985aw} (see, also, \cite{Martin:1986da}), the inequality is not true for any flavor-independent interaction as for some (unrealistic) sharp potentials and mass ratios, the inequality is violated. 

In short, the heavy quarks tend to cluster, benefiting from the maximal chromoelectric interaction. This effect is often  attenuated, when one includes the chromomagnetic interaction, which is maximal for pairs of light quarks. 
\section{Ordering as a function of the number of constituents}
Paradoxically, the quark model was developed in some detail first in the case of baryons, with the pioneering work of Dalitz \cite{1965hep..conf..251D}, while for mesons, it was taken seriously only after the discovery of heavy quarkonia, see, e.g., \cite{Quigg:1979vr}. It is, of course, crucial to understand the link between the meson and the baryon sectors \cite{Lipkin:1978eh,Stanley:1980fe}.

Obviously, a baryon is heavier than a meson made of the same flavor, e.g., in the strange sector:
\begin{equation}\label{eq:mes-bar1}
\phi(s\bar s)< \Omega (sss)~,
\end{equation}
but slightly less trivial is: 
\begin{equation}\label{eq:mes-bar2}
	3\,\phi(s\bar s)< 2\,\Omega (sss)~,
\end{equation}
as the constituent masses cancel out in the balance. Note that \eqref{eq:mes-bar2} and its non-strange analog is well satisfied with e.g., $\phi\sim 1.02\,$GeV and $\Omega\sim 1.67\,$GeV. It means that a quark is on the average heavier in a baryon than in a meson. A variant of \eqref{eq:mes-bar2} reads: 
\begin{equation}\label{eq:mes-bar3}
	3\,(q\bar q)< (qqq)+(\bar q\bar q \bar q)~,
\end{equation}
which implies that the mere rearrangement of three quarks and three antiquarks into three mesons is an allowed process for nucleon-antinucleon annihilation.

Now, the simplest model of the internal baryon dynamics consists of a \emph{pairwise} interaction mediated by a \emph{color-octet} exchange, which reads
\begin{equation}\label{eq:mes-bar4}
V(\vec r_1,\vec r_2,\vec r_3)=\frac12\,\sum_{i<j} v(r_{ij})~,
\end{equation}
where $r_{ij}=|\vec r_j-\vec r_i|$, sometimes dubbed as the ''1/2 rule''. Then, \eqref{eq:mes-bar2} can be easily demonstrated \cite{Ader:1981db,Nussinov:1999sx}from the decomposition
\begin{equation}\label{eq:H3-split}
	H=\frac12\left[\frac{\vec p_1^2}{2\,m}+\frac{\vec p_2^2}{2\,m}+v(r_{12})\right]+\cdots
\end{equation}
that express the equal-mass baryon Hamiltonian as a sum of three meson Hamiltonians. To deduce \eqref{eq:mes-bar3}, it is sufficient to remember that the minimum of a sum is larger that the sum of minima!

Note that the inequality \eqref{eq:mes-bar3} is far from saturation. The reason is that in  the decomposition \eqref{eq:H3-split}, the energy of each bracket is merely bound by its absolute minimum which occurs at rest. In fact, in a three-body system, each two-body pair has its own overall motion. This was noted years ago by Hall and Post, and refined by Basdevant et al.\   For references, see, e.g., the review~\cite{Richard:2019cmi}. There are somewhat overlapping results in this domain, as it deals with different subfields: nuclear physics, stability of matter, hadron spectroscopy, etc. 

Some generalizations to unequal masses are possible \cite{Richard:1984wy,Nussinov:1999sx}, for instance:
\begin{equation}\label{eq:mes-bar5}
(q_1\bar q_2)+(q_2\bar q_3)+(q_3\bar q_1)\le 2\, (q_1q_2q_3)~,
\end{equation}
but the inequality:
\begin{equation}\label{eq:mes-bar6}
	3\,(\overline Q  q)\le (\overline Q{\,}\overline Q{\,}\overline Q)+(qqq)~,
	\end{equation}
that holds for moderate values of the quark mass ratio $M(Q)/m(q)$, but ceases to be valid for large $M/m$. In this latter case, a triply heavy antibaryon would not annihilate into ordinary matter. The critical mass ratio depends on the potential. 

The ``1/2 rule'' relating mesons and baryons can be generalized to multiquarks as:
\begin{equation}
v(r_{ij})=-\frac{3}{16}\ll{i}{j} V_\text{Q}(r_{ij})~,
\end{equation}
where $V_\text{Q}$ denotes the quarkonium potential, and $\tilde \lambda_i$, the eight color operators for the $i^{\rm th}$ quark.  Then, a careful study of the four-body problem shows that the equal-mass tetraquark $qq\bar q\bar q$ is not bound, namely: 
\begin{equation}\label{eq:mult1}
	2\, (q\bar q)\le (qq\bar q\bar q)~,
\end{equation}
As explained in Sec.~\ref{sec:spread}, this result is at variance with respect to the case of atomic physics, where it is established (annihilation is disregarded)  that
\begin{equation}
	(e^+e^+e^-e^-)<2\,(e^+e^-)~. 
\end{equation}

The hierarchy \eqref{eq:mult1} can be inverted for unequal masses, i.e., 
\begin{equation}\label{eq:mult2}
	(QQ\bar q\bar q)\le 2\,(Q\bar q)~,
	\end{equation}
as pointed out years ago~\cite{Ader:1981db,Carlson:1987hh} and reminded in the context of the discovery of the $T_{cc}^+$ state at LHCb~\cite{LHCb:2021vvq}.
\section{Ordering as a function of the spread of couplings}
\label{sec:spread}
The problem of the instability of tetraquark with equal masses, as per \eqref{eq:mult1}
raises the following issue with,
\begin{equation}
	H[\{g\}]=\sum_{i=1}^4 \frac{\vec p_i^2}{2\,m}+\sum_{1\le i<j}^4 g_{ij}\,v(r_{ij})~, \quad \sum_{1\le i<j}^4 g_{ij}=2~,
\end{equation}
as a four-body Hamiltonian of given cumulated strength $\sum g_{ij}=2$, where $v(r)$ is an attractive potential: how does the binding energy evolves when the set of couplings $g_{ij}$ is varied?

For instance, for a set of  two positronium atoms, $g_{ij}=0$ except for $g_{12}=g_{34}=1$, while for the positronium molecule Ps$_2$ and an appropriate numbering, $g_{12}=g_{34}=-1$ and other $g_{ij}=+1$. 
A rigorous result is that if $E[\{g\}]$ denotes the ground state of $H[\{g\}]$
\begin{equation}
	E[\{g\}]\le E[\{\bar g\}]~,
	\end{equation}
where $\{\bar g\}$ denotes the symmetric configuration where $g_{ij}=1/3$ $\forall i,\,j$. This results immediately from the variational principle applied to $H[\{g\}]$ with the wavefunction of $H[\{\bar g\}]$ as a trial function. 

A useful generalization consists of considering: 
\begin{equation}
	\{g\}=\{\bar g\} +\lambda\, \{\tilde g\}~,
	\end{equation}
where $\tilde g_{12}=\tilde g_{34}=-2$ and other $\tilde g_{ij}=-1$. As the real parameter $\lambda$ enters linearly the Hamiltonian, the corresponding ground state energy $E(\lambda)$ is a concave function of $\lambda$. 
Then, as $\lambda=0$ points at the symmetric maximum:
\begin{equation}\label{eq:lambda-sign}
\begin{aligned}
	&0\le \lambda_1\le \lambda_2&\quad \Rightarrow\quad  E(\lambda_2)	\le E(\lambda_1)\le E(0)~,\\
	&\lambda_2\le \lambda_1\le 0&\quad \Rightarrow\quad  E(\lambda_2)	\le E(\lambda_1)\le E(0)~,\\
\end{aligned}
\end{equation}

Now, it can be reasonably expected that $E(\lambda) $ is nearly symmetric around its maximum at $\lambda=0$, leading to the less rigorous result:
\begin{equation}\label{eq:lambda-abs}
	|\lambda_2|>|\lambda_1| \quad \Rightarrow \quad E(\lambda_2)\le E(\lambda_1)~.
\end{equation}
The threshold made of two positronium  atoms corresponds to $\lambda=1/3$, while the positronium molecule Ps$_2$ to $\lambda=-2/3$. According to \eqref{eq:lambda-abs} it makes very plausible the stability of Ps$_2$, which is, indeed, true. Of course, the stability of Ps$_2$ is not taught exactly this way in the textbooks on quantum chemistry, but there is no contradiction: the spread of the coefficients $\{g\}$ in Ps$_2$ explains why a configuration can favor the closeness of the pairs with attraction and the remoteness of the other pairs, and thus polarizes efficiently the two atoms and leads to their binding. 

The above results can also be applied to the equal mass tetraquark with a pure central potential. The color-octet exchange potential for a pure color $\bar 3$-3 configuration corresponds to: 
\begin{equation}
	g_{12}=g_{34}=\frac12~, \quad \text{other} \ g_{ij}=\frac14~
\end{equation}
i.e., $\lambda=1/12\sim 0.083$ associated to a typical interquark potential $v(r)=-a/r+b\,r$, with $a,\,b>0$, while the threshold made of two mesons is still at $\lambda =1/3$. It follows from \eqref{eq:lambda-sign} that this $\bar3$-3 configuration cannot be stable in such a simple chromoelectric model \cite{Richard:2018yrm}. 

For a color 6-$\bar 6$ state, the coefficients are given by: 
\begin{equation}
	g_{12}=g_{34}=-\frac14~, \quad \text{other} \ g_{ij}=\frac58~
\end{equation}
which gives a more favorable value of asymmetry parameter $\lambda=-7/24\sim 0.29$, but still smaller in absolute value than the one of the threshold. Indeed, a pure 6-$\bar 6$ state is not bound.  It is due to  the smaller spread of strength factors, i.e., the non-Abelian  character of the color algebra, that penalizes the tetraquark, as compared to the positronium molecule. 
\section{Outlook}
The Roper resonance has stimulated many discussions about the dynamics of quarks within baryons. Interestingly a more normal ordering $E_{0,1}>E_{1,0}$  is observed in the $\Lambda (uds)$ and $\Lambda_c(udc)$ sectors. Lattice calculations are usually done first with a large pion mass, about 400\,MeV, i.e., for  a large light quark mass $m_{u,d}$, and in the next steps,  the pion mass is decreased. One observes that the degeneracy  $E_{0,1}\simeq E_{1,0}$ shows up only at very small pion mass~\cite{Mathur:2003zf}. It means that the understanding the puzzle requires accounting for the chiral dynamics. Nowadays, the quark model is a little out of fashion, and it is more and more widely  understood that a detailed description of the hadron spectrum requires to account for the coupling to the real or virtual decay channels, either as a correction to the quark model or as a new dynamical scheme. See, e.g.\ \cite{Gavela:1978bz,RevModPhys.91.011003,Golli:2017nid,Owa:2025mep} and refs.\ there.  The paradox is that the physicist of David's generation started with the "bootstrap" theory \cite{Chew:1962mpd}  with all hadrons on the same footing, and then welcomed the quark model as a relief providing a more systematic approach, and now one witnesses a sort of come-back of bootstrap on new grounds. 

\subsection*{Acknowledgments}
I benefited for several decades from discussions about the baryon spectrum with Muhammad Asghar, Jean-Louis Basdevant, Veljko Dmitra\v{s}inovi\'{c}, Sonia Fleck, Marco Genovese, Claude Gignoux, Dieter Gromes, Hallstein H{\o}\-gaa\-sen, Ronald Horgan,  Nathan Isgur, Gabriel Karl, Andr\'e Martin, Willibald Plessas, Michel Fabre de la Ripelle, Winston Roberts, Elena Santopinto,  Bernard Silvestre-Brac, Paul Sorba, Ica Stancu, Pierre Stassart, Alfredo Valcarce, Javier Vijande,  Pierre Taxil and  Sami Zouzou. 
%

\end{document}